\documentclass[review,3p]{elsarticle}

\usepackage[T1]{fontenc} 

\usepackage{graphicx}
\usepackage{hyperref}
\usepackage{amsmath}
\usepackage{breakurl}

\def\figext{eps}
\def\figext{pdf}

\usepackage[mathlines]{lineno}
\usepackage{color}

\usepackage[normalem]{ulem} 
\renewcommand\sout{\bgroup \color[rgb]{0.55,0.00,0.99} \ULdepth=-.5ex \ULset}

\begin{document}

\begin{frontmatter}

\title{NRQCD analysis of charmonium production with pion and proton beams at fixed-target energies}

\author[a,b]{Chia-Yu Hsieh\fnref{myfootnote}}
\fntext[myfootnote]{These authors contributed equally to this work.}

\author[a,c]{Yu-Shiang Lian\fnref{myfootnote}}

\author[a]{Wen-Chen Chang\corref{mycorrespondingauthor}}
\cortext[mycorrespondingauthor]{Corresponding author}
\ead{changwc@phys.sinica.edu.tw}

\author[d]{Jen-Chieh Peng}

\author[e]{Stephane Platchkov}

\author[f]{and Takahiro Sawada}

\address[a]{Institute of Physics, Academia Sinica, Taipei 11529,
  Taiwan}

\address[b]{Department of Physics, National Central University, 300
  Jhongda Road, Jhongli 32001, Taiwan.}

\address[c] {Department of Physics, National Kaohsiung Normal
  University, Kaohsiung County 824, Taiwan}

\address[d]{Department of Physics, University of Illinois at
  Urbana-Champaign, Urbana, Illinois 61801, USA}

\address[e]{IRFU, CEA, Universit\'{e} Paris-Saclay, 91191
  Gif-sur-Yvette, France}

\address[f]{Department of Physics, Osaka City University, Osaka
  558-8585, Japan}


\begin{abstract}
We present an analysis of hadroproduction of $J/\psi$ and $\psi(2S)$
at fixed-target energies in the framework of non-relativistic QCD
(NRQCD). Using both pion- and proton-induced data, a new determination
of the color-octet long-distance matrix elements (LDMEs) is
obtained. Compared with previous results, the contributions from the
$q \bar{q}$ and color-octet processes are significantly enhanced,
especially at lower energies. A good agreement between the
pion-induced $J/\psi$ production data and NRQCD calculations using the
newly obtained LDMEs is achieved. We find that the pion-induced
charmonium production data are sensitive to the gluon density of
pions, and favor pion PDFs with relatively large gluon contents at
large $x$.
\end{abstract}


\end{frontmatter}

\section{Introduction}
\label{sec:introduction}

Since the discovery of the $J/\psi$ resonance in 1974, the study of $c
\bar{c}$ and $b \bar{b}$ quarkonium states has significantly improved
our understanding of the strong interaction~\cite{Schuler:1994hy,
  Schuler:1995sf, Brambilla:2010cs, Andronic:2015wma,
  Lansberg:2019adr}. Production of heavy-quark bound states offers
important testing grounds for both perturbative and nonperturbative
aspects of QCD. From the experimental perspective, quarkonia are
readily detected through their dilepton decay, and the associated
production cross sections are large compared to other hard
processes. Since hadroproduction of quarkonium receives important
contribution form gluon-gluon fusion it could serve as a tool for
accessing the gluon distributions of interacting hadrons, and
particularly for the unstable pion. Our current knowledge on the pion
parton distribution functions (PDFs) mainly comes from pion-induced
Drell-Yan data. As the Drell-Yan cross sections are dominated by
$\bar{q}$-$q$ annihilation, they essentially probe the valence quark
distribution in the pion, but leave the sea and the gluon
distributions unconstrained. The mounting interest in the pion PDFs
is reflected in many publications in the past few years. The pion PDFs
were extracted using new and refined global
analyses~\cite{Barry:2018ort, Novikov:2020snp, Bourrely:2020izp}, or
calculated from new theoretical developments, such as the continuum
Dyson-Schwinger Equations~\cite{Bednar:2018mtf, Cui:2020tdf},
light-front holographic QCD~\cite{deTeramond:2018ecg}, and light-front
Hamiltonian~\cite{Lan:2019vui}. First lattice QCD results on the
$x$-dependence of pion valence structure are also becoming
available~\cite{Chen:2018fwa, Sufian:2019bol, Izubuchi:2019lyk,
  Joo:2019bzr, Sufian:2020vzb, Gao:2020ito}.

A significant amount of pion-induced charmonium data has been
collected in the past. Recently, we investigated the sensitivity of the
fixed-target $J/\psi$ production data to the pion
PDFs~\cite{Chang:2020rdy}. The pion-induced $J/\psi$ data on hydrogen
and several light-mass nuclear targets were compared to
next-to-leading-order (NLO) Color Evaporation Model (CEM)~\cite{CEM}
calculations using the available global-fit parametrizations of the
pion PDFs. Over the energy range considered, the calculations with
pion PDFs determined by SMRS~\cite{Sutton:1991ay},
GRV~\cite{Gluck:1991ey}, JAM~\cite{Barry:2018ort} and
xFitter~\cite{Novikov:2020snp}, are all in reasonable agreement with
the total cross sections data. In contrast, for the longitudinal
momentum $x_F$ distributions, we found that the agreement between data
and calculations strongly depends on the magnitude and shape of the
gluon distribution in the pion. The data favor pion PDFs with larger
gluon strength at large $x$.

To examine a possible model dependence of our previous
observations~\cite{Chang:2020rdy}, we extend our work to a more
elaborate formalism for quarkonia production, namely, the
non-relativistic QCD (NRQCD)~\cite{NRQCD} approach. This approach is
based on a factorization of the heavy-quark pair $Q\bar{Q}$ production
and its subsequent hadronization. The production of the $Q\bar{Q}$
pair proceeds through a short-distance partonic interaction,
calculated using perturbative QCD. The probability of hadronization of
a $Q\bar{Q}$ pair into some quarkonium bound state depends on its
spin, color and angular momentum. Being of a non-perturbative nature,
the hadronization is described by the associated long-distance matrix
elements (LDMEs). The LDMEs, assumed to be universal, are extracted
from the experimental data. The values for the color-singlet (CS)
LDMEs can be obtained from the decay widths of charmonia or model
calculations, while the color-octet (CO) ones are determined from fits
to data of transverse-momentum ($p_T$) differential cross sections,
polarization, and $p_T$-integrated cross sections of charmonium
production.

The NRQCD approach is based on the assumption that the factorization
holds~\cite{NRQCD}. However, for low $p_T$ additional effects might
have large contributions. Analysis of fixed-target data could provide
hints for such effects by assessing the universality of the
LDMEs. Such study, focusing exclusively on charmonium hadroproduction
from fixed-target experiments, was reported by Beneke and
Rothstein~\cite{Beneke:1996tk}. The calculation was performed at
partial next-to-leading order (up to $\mathcal{O}(\alpha_s^3)$). Some
LDMEs were taken from fits to collider data of $p_T$-differential
cross sections~\cite{Eichten:1995ch, Cho:1995ce} while others were
fitted to the proton-induced cross section. The resulting LDMEs were
used to calculate the pion-induced total cross sections. The
calculation was found to lie systematically below the data, for both
$J/\psi$ and $\psi(2S)$ production. The failure in describing the
pion-induced data was speculated to be due to either inaccurate gluon
distributions of pion PDFs or a difference in the higher twist effects
between the proton and the pion data~\cite{Beneke:1996tk}.

More recently, Maltoni \textit{et al.}~\cite{Maltoni:2006yp}
re-analyzed the data of $J/\psi$ and $\psi(2S)$ fixed-target
proton-induced cross sections. The pion-induced data were not included
in the analysis. Their NRQCD calculation was performed at NLO using
the formalism described in Ref.~\cite{Petrelli:1997ge}. The initial CO
parameters were taken from the Tevatron data~\cite{Beneke:1996yw,
  Nason:1999ta}, but the S-wave CO LDMEs for $J/\psi$ and $\psi(2S)$
were allowed to vary. A good description of the data was achieved, at
the expense of a strong reduction of the color-octet LDMEs by a factor
of 5 to 10, depending on the proton PDFs used.

By using the NRQCD formalism in the fixed-target domain, the primary
motivation of our present work is twofold. First, we aim at the
determination of a new set of LDMEs that can reproduce both pion and
proton-induced J/psi production cross sections. As the pion contains a
valence antiquark, the $J/\psi$ cross section for a pion beam has a
larger $q \bar{q}$ contribution than that for a proton beam. Thereby,
the pion-induced data provide a stronger constraint on the LDMEs that
are responsible for the $q \bar{q}$ part of the cross section. Second,
we use this new set of LDMEs to explore the sensitivity of the
pion-induced charmonium data to the parametrizations of the pion
PDFs. The most recent global fits~\cite{Barry:2018ort,Novikov:2020snp}
lead to pion gluon distributions which significantly differ from the
older pion PDF determinations~\cite{Sutton:1991ay, Gluck:1991ey}, at
large and medium parton momenta~\cite{Chang:2020rdy}. According to our
previous results~\cite{Chang:2020rdy}, these new sets of pion PDFs
bring worse agreement between the CEM calculations and the data. It is
important to validate these observations with a NRQCD calculation.

To perform this study, we adopt the NRQCD framework explicitly
formulated in Ref.~\cite{Beneke:1996tk}. We start with a comparison of
NRQCD calculations with the total cross section data for $J/\psi$ and
$\psi(2S)$ productions with proton and pion beams. We then vary the CO
LDMEs for $J/\psi$ and $\psi(2S)$ production to achieve a consistent
description of both proton and pion data. In addition, we use the new
LDMEs to explore the sensitivity of the pion-induced $J/\psi$ data to
the gluon distribution in the pion. This paper is organized as
follows. In Sec.~\ref{sec:NRQCD}, we introduce the NRQCD framework for
the calculations of charmonia cross sections in hadroproduction. A new
fit to the fixed-target $J/\psi$ and $\psi(2S)$ production data to
determine a new set of LDMEs is presented in
Sec.~\ref{sec:results}. We further discuss several findings from the
present study in Sec.~\ref{sec:discussion}, followed by a summary in
Sec.~\ref{sec:summary}.

\section{Charmonium Production and NRQCD}
\label{sec:NRQCD}

In NRQCD, the differential cross section as a function of Feynman $x$
($x_F$), $d\sigma/dx_F$ for a charmonium state $H$ ($H$ = $J/\psi$,
$\psi(2S)$, or $\chi_{cJ}$) from the $hN$, ($h$ = $p$, $\bar{p}$, or
$\pi$) collisions, where $h$ is the beam hadron and $N$ the target
nucleon, is~\cite{Vogt:1999dw}
\begin{align}
\label{eq:eq1}
\frac{d\sigma^{H}}{dx_F}=& \sum\limits_{i,j=q, \bar{q},
  G} \int_{0} ^{1} dx_{1} dx_{2} \delta(x_F - x_1 + x_2) \nonumber \\
 \times f^{h}_{i}(x_1, \mu_{F}) & f^{N}_{j}(x_2, \mu_{F}) \hat{\sigma}[ij \rightarrow H](x_1 P_{h} , x_2 P_{N} , \mu_{F},
\mu_{R}, m_c), \\
\hat{\sigma}[ij \rightarrow H] =&  \sum\limits_{n} C^{ij}_{c \bar{c} [n]} (x_1 P_{h} , x_2 P_{N} , \mu_{F},
\mu_{R}, m_c) \times \langle \mathcal{O}_{n}^{H}[^{2S+1}L_{J}] \rangle
\end{align}
\begin{align}
  x_F = 2 p_L/\sqrt{s} \mbox{,   } x_{1,2} =  \frac{\sqrt{x_F^2+4{M_{c \bar{c}}}^2/s} \pm x_F}{2} 
\end{align}
where $i$ and $j$ label the type of interacting partons (gluons,
quarks and antiquarks), and the $c \bar{c}$ pair is denoted by its
color ($n$), spin ($S$), orbital angular momentum ($L$) and total
angular momentum ($J$). Here $m_c$ and $M_{c \bar{c}}$ are the charm
quark and $c \bar{c}$ pair masses, $f^{h}$ and $f^{N}$ are the
incoming hadron and the target nucleon parton distribution functions,
evaluated at their respective Bjorken-$x$ values, $x_1$ and $x_2$.
The $\mu_F$ and $\mu_R$ are the factorization and renormalization
scales. The total cross sections are obtained by integrating over
$x_F$.

The production cross section $\hat{\sigma}[ij \rightarrow H]$ is the
sum of the products of the short-distance coefficients $C^{ij}_{c
  \bar{c} [n]}$, calculated perturbatively in powers of
$\alpha_s(\mu_R)$, and the nonperturbative parameters, LDMEs $\langle
\mathcal{O}_{n}^{H} [^{2S+1}L_{J}]\rangle$. The coefficients
$C^{ij}_{c \bar{c} [n]}$ describe the production of a $c \bar{c}$ pair
in a specific spin-color state $[n]$, while the parameters $\langle
\mathcal{O}_{n}^{H} [^{2S+1}L_{J}]\rangle$ account for the
hadronization of the $c \bar{c} [n]$ pair into the charmonium state
$H$. The LDMEs are mainly determined by a fit to data and assumed to
be universal, independent of beam or target hadrons and the energy
scale.

In this work, we use the expressions given in
Ref.~\cite{Beneke:1996tk} for computation of $J/\psi$, $\psi(2S)$, and
$\chi_{cJ}$ production via $GG$, $q \bar{q}$ and $qG$ processes. The
scattering processes $q \bar{q} \to Q \bar{Q}$ and $G G \to Q \bar{Q}$
at $\mathcal{O}(\alpha_{s}^2)$ produce $Q \bar{Q}$ pairs in an S-wave
CO state or P-wave CS state. Starting from
$\mathcal{O}(\alpha_{s}^3)$, the S-wave or P-wave CS $Q \bar{Q}$ pairs
can be produced through the scattering processes $G G \to Q \bar{Q} G$
and $q G \to Q \bar{Q} q$~\cite{Cho:1995ce}. We use this NRQCD
framework for calculating the total cross sections of $J/\psi$ and
$\psi(2S)$ production in hadronic collision.

The relationship of LDMEs to the scattering processes for each
charmonium state is summarized in Table~\ref{tab:LDMEproc}. For the $q
\bar{q}$ process, the $c \bar{c}$ pairs are produced at
$\mathcal{O}(\alpha_{s}^2)$ in color octet states, which then hadronize
into various charmonium states with the LDMEs $\langle
\mathcal{O}_{8}^{H}[^{3}S_1] \rangle$. For the $GG$ process, both
$J/\psi$, and $\psi(2S)$ can be produced from either the CO $c
\bar{c}$ at $\mathcal{O}(\alpha_{s}^2)$ or the CS $c \bar{c}$ at
$\mathcal{O}(\alpha_{s}^3)$. The corresponding hadronization LDMEs for
the CO and CS states are $\Delta_8^{H} = \langle
\mathcal{O}_{8}^{H}[^{1}S_0]\rangle + 3/m_{c}^2 \langle
\mathcal{O}_{8}^{H}[^{3}P_0]\rangle + 4/(5m_{c}^2) \langle
\mathcal{O}_{8}^{H}[^{3}P_2]\rangle$ and $\langle
\mathcal{O}_{1}^{H}[^{3}S_1] \rangle$, respectively. The $\chi_{c0}$,
$\chi_{c1}$ and $\chi_{c2}$ can be produced from CS $c \bar{c}$ states
in the $GG$ process at either $\mathcal{O}(\alpha_s^2)$ ($\chi_{c0}$,
$\chi_{c2}$) or $\mathcal{O}(\alpha_s^3)$ ($\chi_{c1}$) with the
corresponding hadronization LDMEs $\langle
\mathcal{O}_{1}^{H}[^{3}P_0] \rangle/m_c^2$, $\langle
\mathcal{O}_{1}^{H}[^{3}P_1] \rangle/m_c^2$ and $\langle
\mathcal{O}_{1}^{H}[^{3}P_2] \rangle/m_c^2$. At
$\mathcal{O}(\alpha_{s}^3)$, the $qG$ process can contribute to the
production of $\chi_{c1}$ via the CS $c \bar{c}$ state which
hadronizes with the LDME $\langle \mathcal{O}_{1}^{H}[^{3}P_1]
\rangle/m_c^2$.

\begin{table}[htbp]   
\centering
\begin{tabular}{|c|c|c|c|}
\hline
 $H$ & $q \bar{q}$ & $GG$ & $qG$ \\
\hline
$J/\psi$, $\psi(2S)$ & $\langle \mathcal{O}_{8}^{H}[^{3}S_{1}] \rangle$ ($\mathcal{O}(\alpha_{s}^2)$) & $ \Delta_8^{H}$$^*$ ($\mathcal{O}(\alpha_{s}^2)$) & \\
 & &  $\langle \mathcal{O}_{1}^{H}[^{3}S_{1}] \rangle$ ($\mathcal{O}(\alpha_{s}^3)$) & \\
\hline
$\chi_{c0}$ & $\langle \mathcal{O}_{8}^{H}[^{3}S_{1}] \rangle$ ($\mathcal{O}(\alpha_{s}^2)$) & $ \langle \mathcal{O}_{1}^{H}[^{3}P_{0}] \rangle$ ($\mathcal{O}(\alpha_{s}^2)$) & \\
\hline
$\chi_{c1}$ & $\langle \mathcal{O}_{8}^{H}[^{3}S_{1}] \rangle$ ($\mathcal{O}(\alpha_{s}^2)$) & $ \langle \mathcal{O}_{1}^{H}[^{3}P_{1}] \rangle$ ($\mathcal{O}(\alpha_{s}^3)$) & $ \langle \mathcal{O}_{1}^{H}[^{3}P_{1}] \rangle$ ($\mathcal{O}(\alpha_{s}^3)$)\\
\hline
$\chi_{c2}$ & $\langle \mathcal{O}_{8}^{H}[^{3}S_{1}] \rangle$ ($\mathcal{O}(\alpha_{s}^2)$) & $ \langle \mathcal{O}_{1}^{H}[^{3}P_{2}] \rangle$ ($\mathcal{O}(\alpha_{s}^2)$) & \\
\hline
\end{tabular}
\caption {Relationship of LDMEs and the associated orders of
  $\alpha_s$ to the scattering processes for various charmonium states
  in the NRQCD framework of Ref.~\cite{Beneke:1996tk}.  $^{*}$:
  $\Delta_8^{H} = \langle \mathcal{O}_{8}^{H}[^{1}S_{0}] \rangle +
  \frac{3}{m_c^2} \langle \mathcal{O}_{8}^{H}[^{3}P_{0}] \rangle +
  \frac{4}{5m_c^2} \langle \mathcal{O}_{8}^{H}[^{3}P_{2}] \rangle $}
\label{tab:LDMEproc}
\end{table}

The number of independent LDMEs is further reduced by applying the
spin symmetry relations~\cite{Beneke:1996tk, Maltoni:2006yp}:
\begin{align}
 \langle \mathcal{O}_{8}^{J/\psi, \psi(2S)} [^{3}P_{J}] \rangle &=
(2J+1) \langle \mathcal{O}_{8}^{J/\psi, \psi(2S)} [^{3}P_{0}] \rangle
\mbox{ for $J=2$} \nonumber \\
\langle \mathcal{O}_{8}^{\chi_{cJ}} [^{3}S_{1}] \rangle &= (2J+1)
\langle \mathcal{O}_{8}^{\chi_{c0}} [^{3}S_{1}] \rangle \mbox{ for
  $J=1, 2$} \nonumber \\
\langle \mathcal{O}_{1}^{\chi_{cJ}} [^{3}P_{J}] \rangle &= (2J+1)
\langle \mathcal{O}_{1}^{\chi_{c0}} [^{3}P_{0}] \rangle \mbox{ for
  $J=1, 2$}.
\end{align}

The best-fit LDMEs for the description of proton-induced $J/\psi$ and
$\psi(2S)$ production from Ref.~\cite{Beneke:1996tk} are summarized in
Table~\ref{tab:LDME}. In Ref.~\cite{Beneke:1996tk}, the CS LDMEs
$\langle \mathcal{O}_{1}^{H}[^{3}S_{1}] \rangle$ for $J/\psi$ and
$\psi(2S)$, and $\langle \mathcal{O}_{1}^{H}[^{3}P_{0}] \rangle$ for
$\chi_{c0}$, are given by the potential models~\cite{Eichten:1995ch},
and the CO LDMEs $\langle \mathcal{O}_{8}^{H}[^{3}S_{1}] \rangle$ for
$J/\psi$, $\psi(2S)$ and $\chi_c$, are taken from the fits to Tevatron
collider data of $p_T$ spectra~\cite{Cho:1995ce}. The $\Delta_8^{H}$
parameters for the individual $J/\psi$ and $\psi(2S)$ production were
determined by a fit of NRQCD calculation to the proton-induced
data.

A later study by Maltoni \textit{et al.}~\cite{Maltoni:2006yp} was
performed with a full NLO calculation~\cite{Petrelli:1997ge}. Starting
with the LDMEs determined from the collider data of $p_T$ differential
cross sections~\cite{Beneke:1996yw, Nason:1999ta}, an additional
reduction factor of about 0.1 for $\langle
\mathcal{O}_{8}^{H}[^{3}S_{1}] \rangle$ and $\langle
\mathcal{O}_{8}^{H}[^{1}S_{0}] \rangle$ of $J/\psi$ and $\psi(2S)$,
was required in the calculations to achieve a good description of
fixed-target proton-induced data.

The CS LDMEs used in these two studies are either identical or very
similar, as they were obtained from a potential
model~\cite{Eichten:1995ch}. The CO $ \langle
\mathcal{O}_{8}^{H}[^{3}S_{1}] \rangle$ for $\chi_{c0}$ are nearly
identical. In contrast, the remaining CO LDMEs obtained in
Ref.~\cite{Beneke:1996tk} are about a factor of 1.5-5 larger than
those determined in Ref.~\cite{Maltoni:2006yp}.

\begin{table}[htbp]   
\centering
\begin{tabular}{|c|c|c|c|c|}
\hline
$H$ & $ \langle \mathcal{O}_{1}^{H}[^{3}S_{1}] \rangle$ & $ \langle \mathcal{O}_{1}^{H}[^{3}P_{0}] \rangle/{m_c}^2$ &  $ \langle \mathcal{O}_{8}^{H}[^{3}S_{1}] \rangle$ & $ \Delta_8^{H}$ \\
\hline
$J/\psi$ & $1.16$ &  &$6.6 \times 10^{-3}$ & $3 \times 10^{-2}$ \\
$\psi(2S)$ & $0.76$ &  & $4.6 \times 10^{-3}$ & $5.2 \times 10^{-3}$ \\
$\chi_{c0}$ & & $0.044$ & $3.2 \times 10^{-3}$ & \\
\hline
\end{tabular}
\caption {NRQCD LDMEs for the charmonium production in
  Ref.~\cite{Beneke:1996tk}, all in units of $\rm{GeV}^3$.}
\label{tab:LDME}
\end{table}

With the information of LDMEs, the direct production cross sections of
$J/\psi$, $\psi(2S)$ and three $\chi_{cJ}$ states can be calculated as
shown in Eq.(\ref{eq:eq1}). Furthermore, taking into account the
direct production of $J/\psi$ and the feed-down from hadronic decays
of $\psi(2S)$ and radiative decays of three $\chi_{cJ}$ states, the
total $J/\psi$ cross section is estimated as follows,
\begin{equation}
\sigma_{J/\psi} = \sigma_{J/\psi}^{direct} + Br(\psi(2S) \rightarrow
J/\psi X) \sigma_{\psi(2S)} + \sum \limits_{J=0}^{2} Br( \chi_{cJ}
\rightarrow J/\psi \gamma)\sigma_{\chi_{cJ}}
\end{equation}
For the branching ratios we take the PDG 2020 values~\cite{PDG2020}
\begin{align}
Br(\psi(2S) \rightarrow J/\psi X) & = 61.4 \%, & Br(\chi_{c0} \rightarrow J/\psi \gamma) & = 1.4 \% \nonumber \\
Br(\chi_{c1} \rightarrow J/\psi \gamma) & = 34.3 \%, & Br(\chi_{c2} \rightarrow J/\psi \gamma) & = 19.0 \%.
\end{align}

Using this NRQCD framework, we calculate below the $J/\psi$ and
$\psi(2S)$ production cross sections and compare them to the available
fixed-target data. We explore if a set of LDMEs can be identified to
achieve a good description of both proton-induced and pion-induced
data of charmonia production.

\section{Results of NRQCD calculations}
\label{sec:results}

In this section, we compare our NRQCD calculations with the
fixed-target data of $J/\psi$ and $\psi(2S)$ production by proton and
pion beams. The cross sections for proton-induced $J/\psi$ and $\psi
(2S)$ and their ratios $R_{\psi}=\sigma (\psi (2S))/\sigma (J/\psi)$
are taken from Ref.~\cite{Maltoni:2006yp}. The pion-induced data for
$J/\psi$ and $\psi (2S)$ are taken from Refs.~\cite{Schuler:1994hy,
  Vogt:1999cu}. In addition, three measurements of $R_{\psi}$ from
HERA~\cite{Abt:2006va} and NA38~\cite{Baglin:1994ui}, and the
pion-induced data of $\sigma (J/\psi)$ and $\sigma (\psi (2S))$ from
NA38~\cite{Corden:1980rb} and WA92~\cite{Alexandrov:1999ch} are added.

In spite of a caveat of double counting by including the data of
$R_{\psi}$ in the analysis~\cite{Maltoni:2006yp}, the cancellation of
systematic uncertainties in this quantity brings additional
constraining power in the fit. The NRQCD calculations presented in
this section are performed using the nucleon CT14nlo
PDFs~\cite{CT14nlo} and the pion GRV NLO PDFs~\cite{Gluck:1991ey}
under the LHAPDF framework~\cite{LHAPDF5, LHAPDF6}. The cross sections
are evaluated with a charm quark mass $m_c= 1.5$ GeV/$c^2$ and
renormalization and factorization scale $\mu_R = \mu_F = 2
m_c$~\cite{Beneke:1996tk}.

\begin{figure}[htbp]
\centering
\includegraphics[width=0.93\columnwidth]{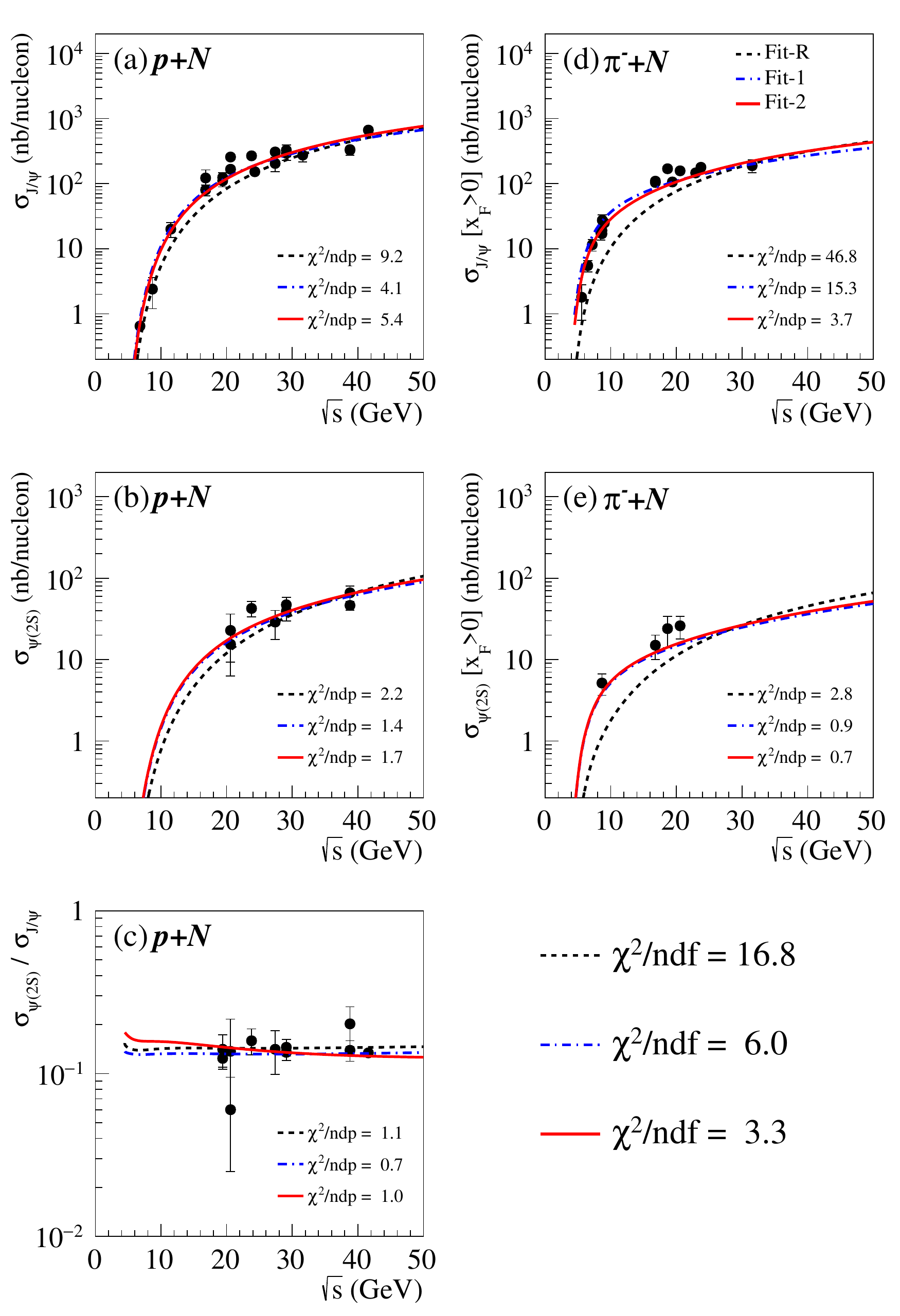}
\caption[\protect{}]{$J/\psi$ and $\psi(2S)$ production cross sections
  and the $\psi(2S)/(J/\psi)$ ratios in the $p+N$ reactions, and
  $J/\psi$ and $\psi(2S)$ production cross sections in the $\pi^- +N$
  reactions, labeled as (a)-(e) in the plot. The dashed (black),
  dot-dashed (blue) and solid (red) curves represent the NRQCD results
  using the LDMEs obtained in "Fit-R", "Fit-1" and "Fit-2",
  respectively. The reduced $\chi^2/\text{ndf}$ for all data are displayed in
  the bottom-right. The values of $\chi^2$ divided by the number of
  data point (ndp) for each data set are also shown.}
\label{fig_LDME}
\end{figure}

\subsection{Reference fit}
\label{sec:results1}

We compare these data with the NRQCD calculations using the LDMEs
determined in Ref.~\cite{Beneke:1996tk}, named as "Fit-R" in the
following. The difference between the proton PDFs selected for the study of
Ref.~\cite{Beneke:1996tk} (CTEQ3L) and here (CT14nlo) has a negligible
effect on the final results. The results are plotted as black dashed
lines in Fig.~\ref{fig_LDME}. The total reduced $\chi^2/\text{ndf}$, together
with the deviations between data and calculations per data point
($\chi^2/\text{ndp}$) for each data set, are displayed in Fig.~\ref{fig_LDME}
and listed in Table~\ref{tab:LDME_FIT}. The reduced $\chi^2/\text{ndf}$ of
the entire data sets is 16.8. Clearly, the pion-induced data below
$\sqrt{s}=20$ GeV are significantly underestimated by the calculations
as already observed in Ref.~\cite{Beneke:1996tk}.

\begin{table}[htbp]   
\centering
\begin{tabular}{|c|c|c|c|}
\hline
 & Fit-R & Fit-1 & Fit-2 \\
\hline
 $\chi^2_{total}/\text{ndf}$  &  16.8 &   6.0 &   3.3 \\ 
\hline
 $\chi^2/\text{ndp}|^{p}_{\sigma(J/\psi)}$ &   9.2 &   4.1 &   5.4 \\ 
 $\chi^2/\text{ndp}|^{p}_{\sigma(\psi(2S))}$ &   2.2 &   1.4 &   1.7 \\ 
 $\chi^2/\text{ndp}|^{p}_{R(\psi(2S))}$ &   1.1 &   0.7 &   1.0 \\ 
\hline
 $\chi^2/\text{ndp}|^{\pi^-}_{\sigma(J/\psi)}$ &  46.8 &  15.3 &   3.7 \\ 
 $\chi^2/\text{ndp}|^{\pi^-}_{\sigma(\psi(2S))}$ &   2.8 &   0.9 &   0.7 \\ 
\hline
 $ \langle \mathcal{O}_{8}^{J/\psi}[^{3}S_{1}] \rangle$ & $6.6 \times 10^{-3}$ & $ (1.47 \pm 0.07) \times 10^{-1}$ & $ (9.5 \pm 0.4) \times 10^{-2}$ \\ 
 $ \Delta_8^{J/\psi}$ & $3 \times 10^{-2}$ & $(0 \pm 8) \times 10^{-4}$ & $(1.8 \pm 0.2) \times 10^{-2}$ \\ 
 $ \langle \mathcal{O}_{8}^{\psi(2S)}[^{3}S_{1}] \rangle$ & $4.6 \times 10^{-3}$ & $(2.5 \pm 0.2) \times 10^{-2}$ & $(2.6 \pm 0.2) \times 10^{-2}$ \\ 
 $ \Delta_8^{\psi(2S)}$ & $5.2 \times 10^{-3}$ & $(0 \pm 8) \times 10^{-4}$ & $(4 \pm 6) \times 10^{-4}$ \\ 
\hline
\end{tabular}
\caption {The reduced $\chi^2/\text{ndf}$ of the whole data sets, the
  $\chi^2$ divided by the number of data point (ndp) for each data set
  in "Fit-R", "Fit-1" and "Fit-2" NRQCD calculations and the
  corresponding input or best-fit LDMEs. All LDMEs are in units of
  $\rm{GeV}^3$.}
\label{tab:LDME_FIT}
\end{table}

\subsection{Fit of the proton-induced data}
\label{sec:results2}

It is known that the $q \bar{q}$ contribution plays an important role
in the pion-induced $J/\psi$ production near threshold because the
parton density at large $x$ is dominated by the valence antiquarks in
pions~\cite{Chang:2020rdy}. In our NRQCD framework, the $\langle
\mathcal{O}_{8}^{H}[^{3}S_{1}] \rangle$ LDMEs are responsible for the
$q \bar{q}$ contribution, as shown in Table~\ref{tab:LDMEproc}, and
their values were taken from the fits to the Tevatron
data~\cite{Cho:1995ce} at high energies. The observed underestimation
of low-energy pion-induced data in "Fit-R" could arise from too small
a value for the input $\langle \mathcal{O}_{8}^{H}[^{3}S_{1}]
\rangle$.

To seek an improved description of the pion-induced data, we take a
different approach, called "Fit-1", of leaving both $\langle
\mathcal{O}_{8}^{H}[^{3}S_{1}] \rangle$ and $ \Delta_8^{H}$ LDMEs for
$J/\psi$ and $\psi(2S)$ as free parameters in the global fit of the
proton-induced data. The CO LDME of $\chi_{c0}$ is fixed at the value
given in Ref.~\cite{Beneke:1996tk} and the best-fit LDMEs are required
to be positive-definite. The resulting fit is shown as blue dot-dashed
lines in Fig.~\ref{fig_LDME}. The cross sections for both proton- and
pion-induced data are significantly enhanced, compared to the results
of "Fit-R". The agreement between the data and calculation is greatly
improved. As shown in Table~\ref{tab:LDME_FIT}, the overall
$\chi^2/\text{ndf}$ is reduced from 16.8 to 6.0, compared to "Fit-R". While
the pion-induced data are not used in the global fit for the LDMEs
determination, these data are included in the evaluation of
$\chi^2/\text{ndf}$ for comparison purposes.

More specifically, the agreement between the pion-induced $J/\psi$
data alone and the calculations with best-fit LDMEs of "Fit-1" is
improved by a factor of 3, compared to "Fit-R". The values of newly
determined $\langle \mathcal{O}_{8}^{H}[^{3}S_{1}] \rangle$ LDMEs are $1.5
\times 10^{-1}$ and $2.5 \times 10^{-2}$ for $J/\psi$ and $\psi(2S)$,
respectively. Both are significantly larger than the "Fit-R" values of
$6.6 \times 10^{-3}$ and $4.6 \times 10^{-3}$ determined from collider
data. The increase of the values of $\langle
\mathcal{O}_{8}^{H}[^{3}S_{1}] \rangle$ clearly accounts for better
agreement between the NRQCD calculation and the pion-induced data, even
though the pion data were not included in the fit.

The values of the CO $ \Delta_8^{H}$ LDMEs for $J/\psi$ and $\psi(2S)$
resulting from "Fit-1" are compatible with zero, as shown in
Table~\ref{tab:LDME_FIT}. Despite an improved description of data in
this approach, "Fit-1" finds vanishing values of $ \Delta_8^{H}$ for
both $J/\psi$ and $\psi(2S)$. It appears that these LDMEs cannot be
determined from the proton-induced $J/\psi$ and $\psi(2S)$ production
data alone. This suggests the need to include also the pion data in
the global fit, as discussed next.

\subsection{Fit of both the pion- and proton-induced data}
\label{sec:results3}

Because of the different nature of valence quarks in the protons and
pions, the energy dependence of the relative contributions of $q
\bar{q}$ and $GG$ processes is different for the $J/\psi$ and
$\psi(2S)$ production, especially at low energies. Under the
assumption that higher-twist effects are negligible, a combined fit of
these two data sets should further constrain the CO LDMEs. The results
of this new fit, referred to as "Fit-2", are shown in
Table~\ref{tab:LDME_FIT} and displayed as the solid red lines in
Fig.~\ref{fig_LDME}.

Comparing the results of "Fit-2" and "Fit-1", it is found that the
description of the pion-induced data is improved, while maintaining
a good agreement between the proton data and the calculation. The
CO matrix elements are also better constrained. The total reduced
$\chi^2/\text{ndf}$ is further decreased to about 3.3. The agreement between
the pion-induced $J/\psi$ data and the NRQCD calculation is improved
by a factor of 4, from a $\chi^2/\text{ndp}$ of 15.3 in "Fit-1" to 3.7 in
"Fit-2". The values of the newly determined CO $\langle
\mathcal{O}_{8}^{H}[^{3}S_{1}] \rangle$ LDMEs are $9.5 \times 10^{-2}$
and $2.6 \times 10^{-2}$ for $J/\psi$ and $\psi(2S)$, respectively,
either smaller than or consistent with those obtained in "Fit-1".

With the inclusion of the pion-induced data, non-zero values of the CO
$ \Delta_8^{H}$ LDMEs can now be obtained. As shown in
Table~\ref{tab:LDME_FIT}, the values of CO $ \Delta_8^{H}$ LDMEs are
found to be $1.8 \times 10^{-2}$ and $4.0 \times 10^{-4}$ for $J/\psi$
and $\psi(2S)$, respectively. The best-fit $\langle
\mathcal{O}_{8}^{H}[^{3}S_{1}] \rangle$ LDMEs responsible for the
contribution of the $q \bar{q}$ process are larger by about a factor
of 10, while the $ \Delta_8^{H}$ related to the contribution of the
$GG$ process are reduced by a factor of 2-10, in comparison with the
LDMEs determined from collider data~\cite{Cho:1995ce,
  Beneke:1996yw}. The new CO LDMEs indicate that the $q \bar{q}$
contribution determined by the fixed-target data is significantly
larger than the corresponding contribution at collider energies.


\begin{figure}[htbp]
\centering
\includegraphics[width=1.0\columnwidth]{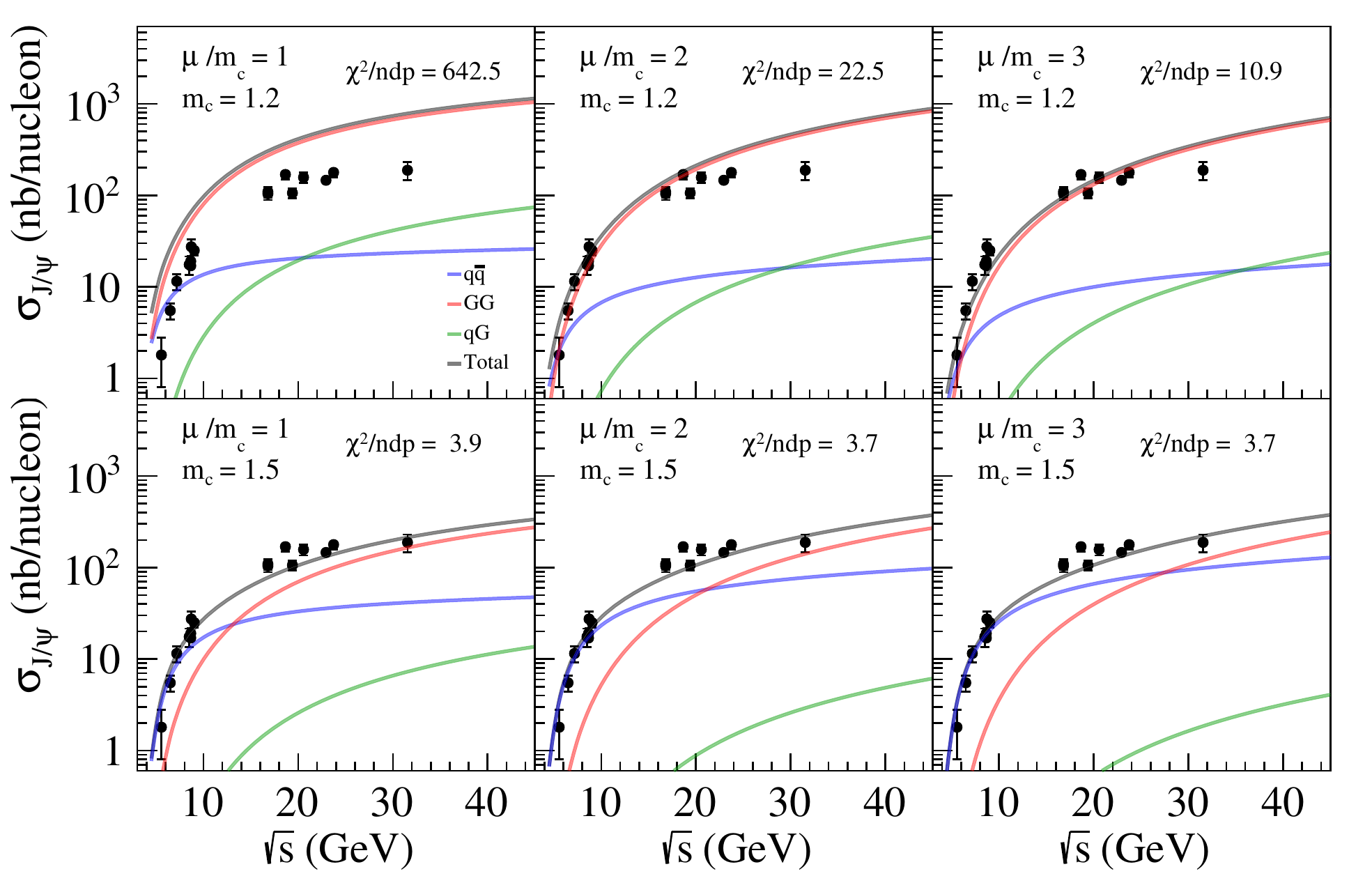}
\caption[\protect{}]{The NRQCD results from an analysis of both
  proton-induced and pion-induced data sets under variation of charm
  quark mass $m_c$, renormalization scale $\mu_R$ and factorization
  scale $\mu_F$, compared with the pion-induced data of $J/\psi$
  production as a function of $\sqrt{s}$. The total cross sections and
  $q \bar{q}$, $GG$, and $qG$ contributions are denoted as black,
  blue, red and green lines, respectively. The values of $m_c$,
  $\mu=\mu_R=\mu_F$ in the NRQCD calculation as well as the best-fit
  $\chi^2/\text{ndp}$ are displayed in each plot.}
\label{fig_sys_GRV}
\end{figure}

The systematic uncertainties of these results are studied by setting
$m_c$ to 1.2 GeV/$c^2$ or 1.5 GeV/$c^2$ and the normalization scale
$\mu = \mu_R = \mu_R$ to $m_c$, 2$m_c$, and
3$m_c$. Figure~\ref{fig_sys_GRV} shows the comparison of the
pion-induced $J/\psi$ data and NRQCD calculation with the
corresponding settings of $m_c$ and $\mu$. The total cross sections
and the $q \bar{q}$, $GG$, and $qG$ contributions are denoted as
black, blue, red and green lines, respectively.

The parameter of the charm quark mass $m_c$ plays a significant role in
the systematic effect. With $m_c$ set to 1.2 GeV/$c^2$, the CO LDMEs
as free parameters are not well constrained, and the quality of fit
significantly deteriorates as seen from the increased
$\chi^2/\text{ndp}$. Judging from the contributions of various subprocesses,
the $GG$ process is enhanced too much to provide a good description of
data in the calculations with this reduction of $m_c$. When $m_c$ is
set as 1.5 GeV/$c^2$, the quality of the fit is equally good with
$\mu$ varying among $m_c$, 2$m_c$, and 3$m_c$. Even though one of CO
LDMEs cannot be obtained with good accuracy when $\mu$ is set at
$m_c$, the best-fit LDMEs for three different scales are consistent
with having large values of $\langle \mathcal{O}_{8}^{H}[^{3}S_{1}]
\rangle$ LDMEs, reflecting a non-negligible $q \bar{q}$ contribution.

\section{Discussion}
\label{sec:discussion}

\subsection{Fractions of individual contributions in hadroproduction}

A new set of LDMEs for $J/\psi$ and $\psi(2S)$ production at
fixed-target energies has been obtained in an analysis of data with
proton and pion beams. Our analysis differs from that of
Ref.~\cite{Beneke:1996tk} in two aspects. First, the LDMEs $\langle
\mathcal{O}_{8}^{H}[^{3}S_{1}] \rangle$ are now allowed to vary in the
global fit. This leads to a much improved description of fixed-target
data with proton beam. Second, the pion data are included in the
global fit. This allows for the determination of the LDMEs
$\Delta_8^{H}$.

To better understand the reasons for the significantly improved
description of the $J/\psi$ and $\psi(2S)$ production data, it is
instructive to compare the NRQCD calculations using the "Fit-R" LDMEs
of Ref.~\cite{Beneke:1996tk} and the "Fit-2" LDMEs of the present
analysis. In particular, we examine the decomposition of the $J/\psi$
production cross section into individual contributions in three
fashions: (i) $q \bar{q}$, $GG$, and $qG$ subprocesses; (ii) color
singlet versus color octet $c \bar{c}$ states; (iii) direct production
of $J/\psi$ versus feed-down from $\psi(2S)$ and $\chi_{c}$.


\begin{figure}[htbp]
\centering
{\includegraphics[width=1.0\columnwidth]{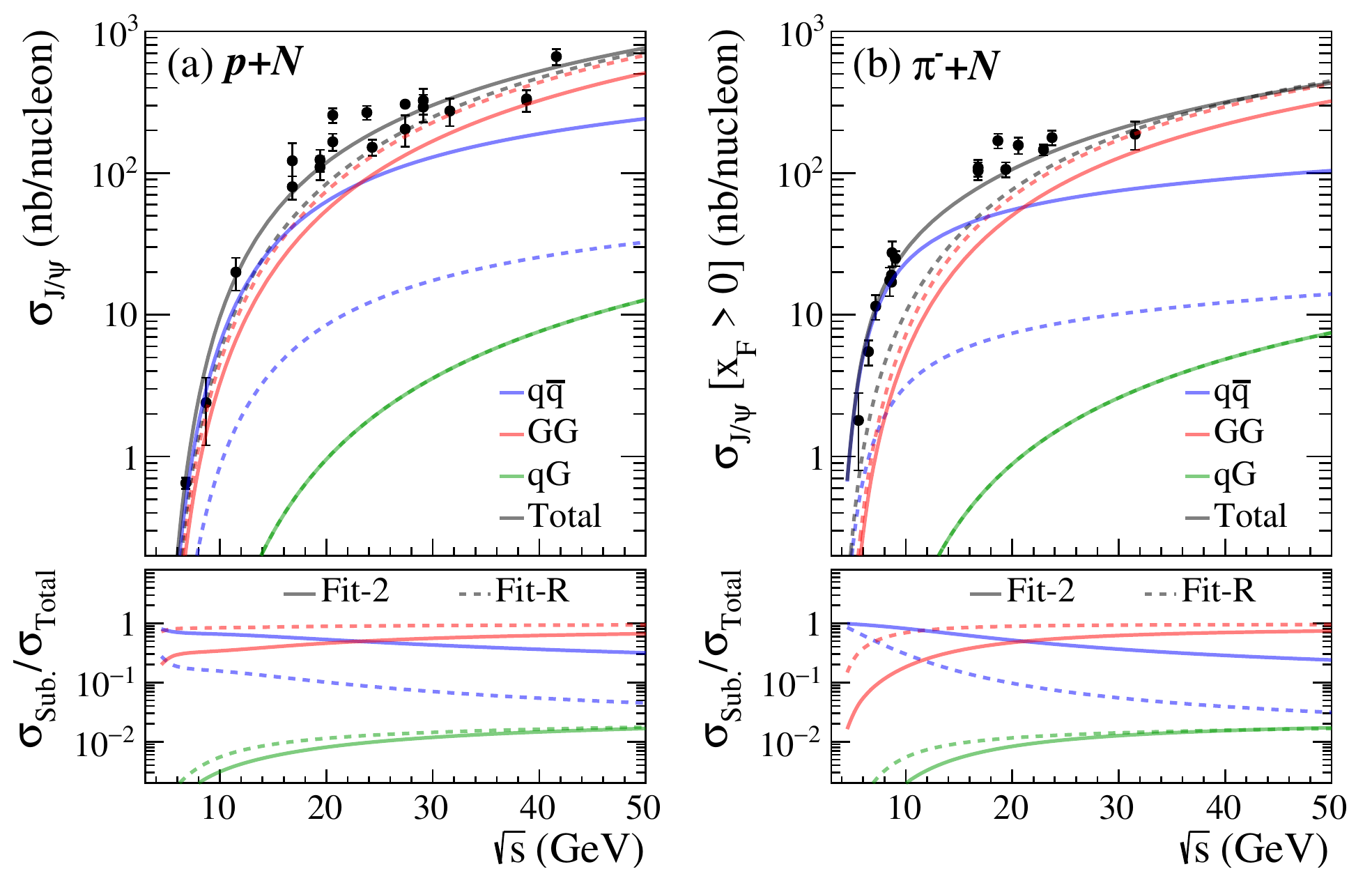}}
\caption
[\protect{}]{Total cross sections (black) and contributions from $q
  \bar{q}$ (blue), $GG$ (red) and $qG$ (green) processes for $J/\psi$
  production as a function of $\sqrt{s}$ in (a) $pN$ and (b) $\pi^- N$
  interactions. The dashed and solid curves represent the "Fit-R" and
  "Fit-2" results. The fractions of each sub-process cross section are
  displayed at the bottom of each plot.}
\label{fig_jpsi_view1}
\end{figure}

The decomposition of $J/\psi$ production cross sections for proton and
pion beams into the $q \bar{q}$, $GG$ and $qG$ processes is shown in
Fig.~\ref{fig_jpsi_view1}. We note that the $qG$ contributions remain
unchanged in the new analysis since the $qG$ process only contributes
to the formation of the $\chi_{c1}$ states and the LDMEs for
$\chi_{cJ}$ are identical for "Fit-R" and
"Fit-2". Figure~\ref{fig_jpsi_view1} also shows that the $GG$
contribution is dominant in the $J/\psi$ production with proton beam
at all energies, except near the threshold. In contrast, the $q
\bar{q}$ contribution for pion-induced data is enhanced due to the
increased antiquark content in pion's valence region. Therefore, the
inclusion of the pion data in the global fit provides additional
constraints on those LDMEs which are sensitive to the $q \bar q$
process. The low-energy fixed-target pion data are particularly
important for the determination of the $\langle
\mathcal{O}_{8}^{H}[^{3}S_{1}] \rangle$ LDMEs.

In comparison with "Fit-R", the $\langle
\mathcal{O}_{8}^{H}[^{3}S_{1}] \rangle$ LDME is increased, whereas $
\Delta_8^{H}$ is decreased. These changes lead to an enhancement of
the CO $q \bar{q}$ contribution and a reduction of the CO $GG$
contribution. The increase of the fraction of $q \bar{q}$
contribution, especially at low-energies, accounts for the improvement
in describing the pion data.  The opposite trend in the variations of
the two CO LDMEs $\langle \mathcal{O}_{8}^{H}[^{3}S_{1}] \rangle$ and
$ \Delta_8^{H}$ leads to significant changes in the energy dependence
of $J/\psi$ production cross sections, as shown in
Fig.~\ref{fig_jpsi_view1}.

\begin{figure}[htbp]
\centering
{\includegraphics[width=1.0\columnwidth]{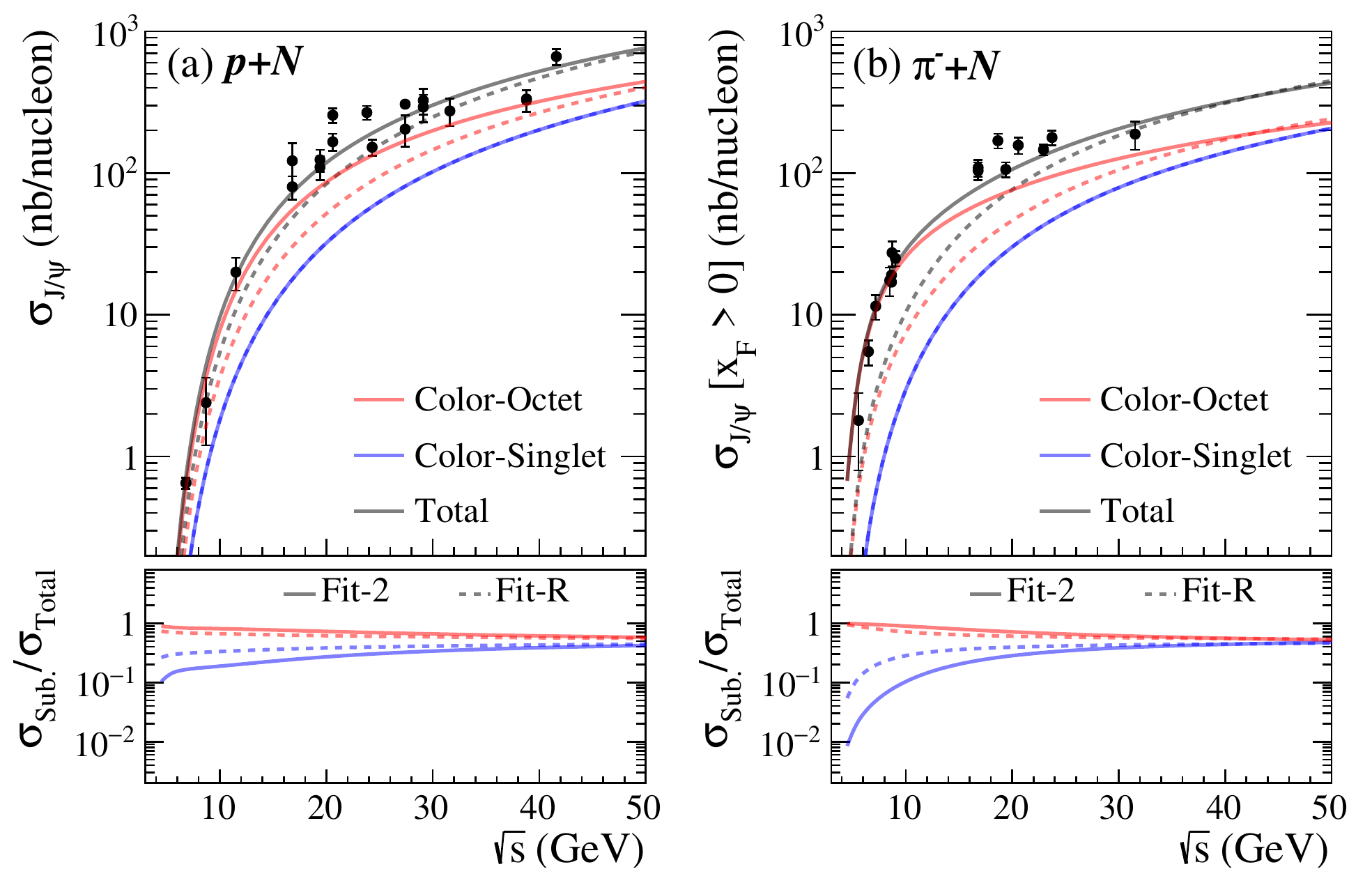}}
\caption
[\protect{}]{Same as Fig.~\ref{fig_jpsi_view1} for the decomposition
  of contributions from CS (blue) and CO (red) processes.}
\label{fig_jpsi_view2}
\end{figure}

Figure~\ref{fig_jpsi_view2} shows the decomposition of the $J/\psi$
cross sections into contributions from the color octet and color
singlet states. As the CS contribution (blue lines) in our study is
fixed, the enlarged $\langle \mathcal{O}_{8}^{H}[^{3}S_{1}] \rangle$
LDMEs significantly enhance the CO contribution (red lines) at low
energies, while in "Fit-2" the reduced $\Delta_8^{H}$ results in a
reduction of the CO contribution at high energies. Through the interplay
of these two CO LDMEs and the parton luminosities, the CO contribution
remains similar at high energies but is enhanced near threshold for
the proton-induced production. In the case of pion-induced production,
the CO contribution is slightly suppressed at high energies.

\begin{figure}[htbp]
\centering
{\includegraphics[width=1.0\columnwidth]{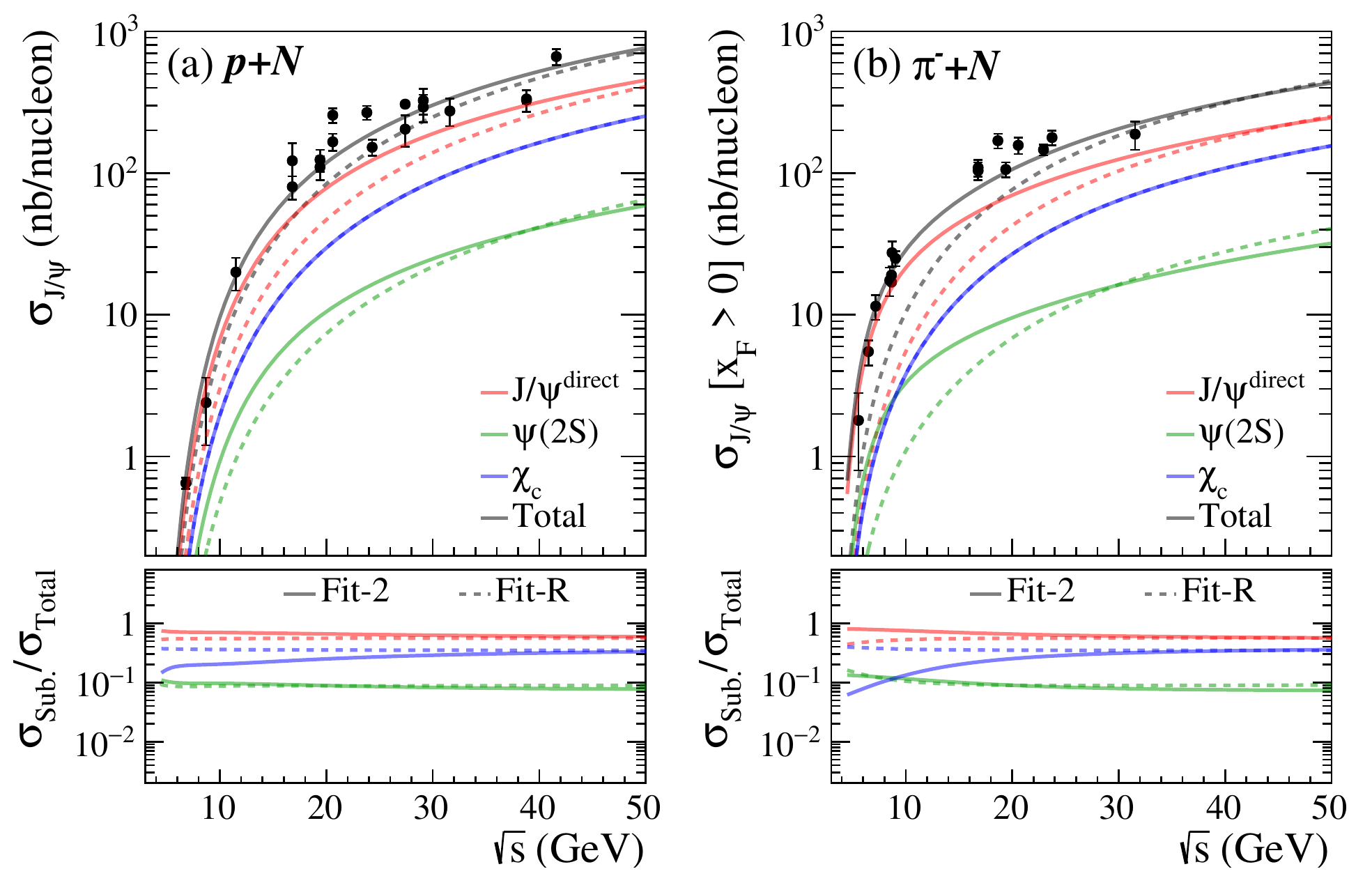}}
\caption
[\protect{}]{Same as Fig.~\ref{fig_jpsi_view1} for the decomposition
  of contributions of $J/\psi$ production from direct production of
  $J/\psi$ (red) and feed-down from $\psi(2S)$ (green) and all $\chi_c$
  states (blue).}
\label{fig_jpsi_view3}
\end{figure}

Figure~\ref{fig_jpsi_view3} shows the decomposition of the $J/\psi$
cross sections into the contributions from direct production (red
lines) and the feed-down from heavier charmonium states of $\psi(2S)$
(blue lines) and $\chi_c$ (green lines). We note that the LDMEs for
the three $\chi_c$ states are kept unchanged for "Fit-R" and
"Fit-2". The most notable change between the calculations with "Fit-2"
and the "Fit-R" is the enhancement of the direct $J/\psi$ production
at low energies, as a consequence of an enlarged $\langle
\mathcal{O}_{8}^{H}[^{3}S_{1}] \rangle$. Taking into account the decay
branching ratios, the contributions to the $J/\psi$ production in
descending order of importance are direct production, $\chi_{c}$, and
$\psi(2S)$.

To recapitulate the main findings at this point, we note that the
inclusion of the low-energy pion-induced total cross section data of
$J/\psi$ and $\psi(2S)$ production to the analysis of NRQCD provides
an important constraint of the $\langle \mathcal{O}_{8}^{H}[^{3}S_{1}]
\rangle$ LDMEs via the $q \bar{q}$ contributions. A good description
of both proton- and pion-induced data by NRQCD can be achieved. The $q
\bar{q}$ and CO contributions from the NRQCD calculations with the new
LDMEs are greatly enhanced at low energies with proton and pion beams,
compared with results found in the earlier
studies~\cite{Beneke:1996tk, Maltoni:2006yp}.

\subsection{Sensitivity to the pion PDFs}

The new set of LDMEs is now used to study the sensitivity of the
$J/\psi$ data to the various pion PDFs.  We have considered four pion
PDFs, namely, SMRS~\cite{Sutton:1991ay} and GRV~\cite{Gluck:1991ey},
representative of the most widely used pion PDFs thus far, as well as
JAM~\cite{Barry:2018ort} and xFitter~\cite{Novikov:2020snp}, obtained
from very recent global analyses. For SMRS, we select the default one
in which the sea quarks carry 15\% of the pion momentum at $Q^2$= 4
GeV$^2$. As illustrated in Ref.~\cite{Chang:2020rdy}, SMRS, JAM, and
xFitter have similar valence-quark distributions, while the magnitude
of the GRV distribution is smaller by about 20\%--30\%. For the gluon
distributions, SMRS and GRV have similar shapes and magnitudes, while
xFitter and JAM have significantly smaller magnitudes by a factor of
2--4.

\begin{figure}[htbp]
\centering
{\includegraphics[width=1.0\columnwidth]{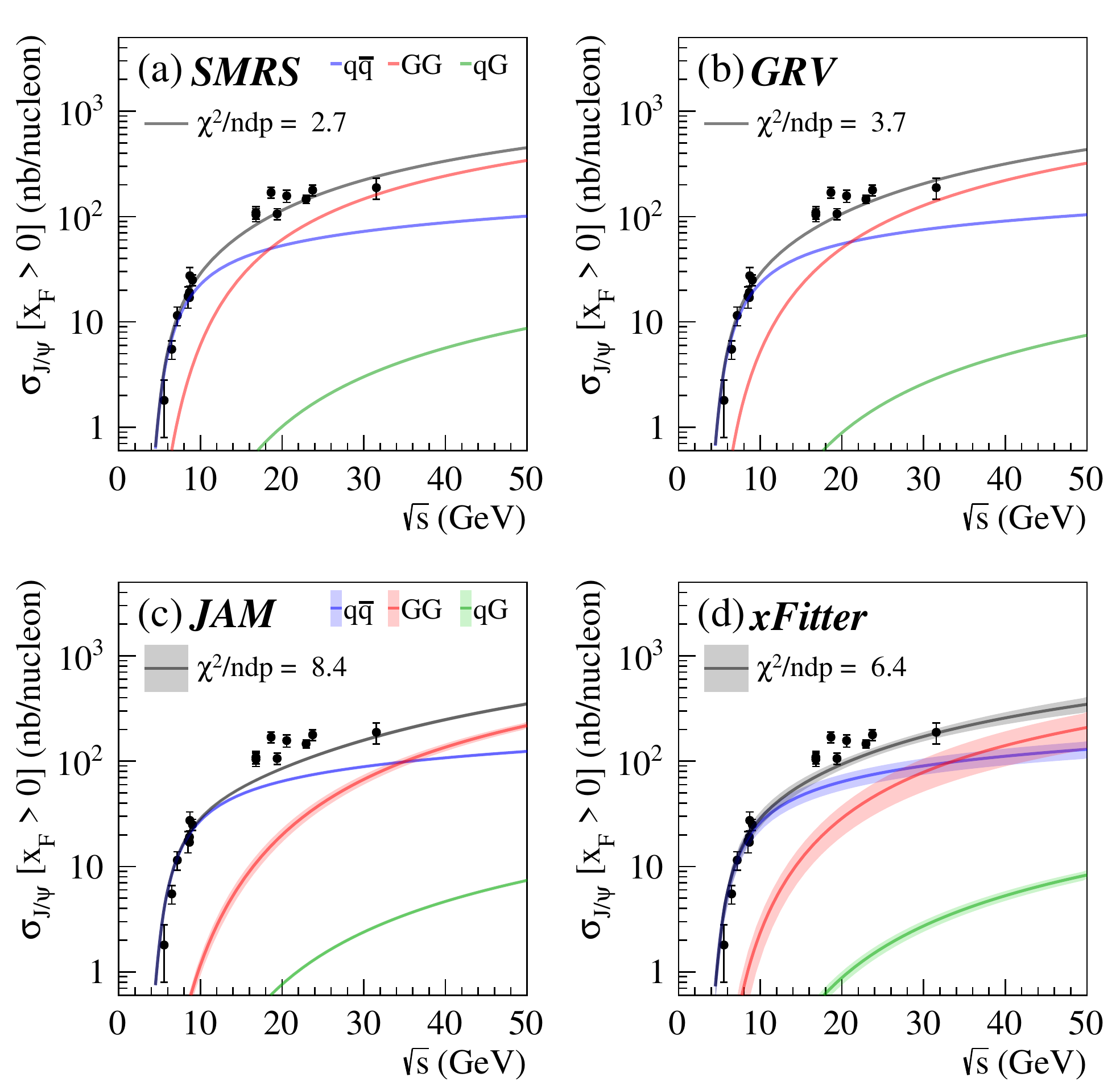}}
\caption
[\protect{}] {The NRQCD $J/\psi$ production cross sections at $x_F >0$
  for the $\pi^- N$ reactions, calculated for four pion PDFs (SMRS,
  GRV, JAM and xFitter) using LDMEs of "Fit-2". The black, blue, red,
  and green curves represent the calculated total cross sections, and
  the $q \bar{q}$, $GG$, and $qG$ contributions, respectively. The
  shaded bands on the xFitter and JAM calculations come from the
  uncertainties of the corresponding PDF sets. The SMRS and GRV PDFs
  contain no information on uncertainties.}
\label{fig_jpsi_PDF}
\end{figure}

The NRQCD calculation with each of the four pion PDFs is compared with
the data in Fig.~\ref{fig_jpsi_PDF}. Overall, the total cross sections
(black lines) for the four pion PDFs exhibit similar $\sqrt{s}$
dependencies. However, the individual terms differ strongly. The $q
\bar{q}$ contribution dominates near thresholds and the $GG$
contribution increases rapidly at higher energies, while the $qG$
component is relatively negligible over the whole energy range. The
relative fractions of $q \bar{q}$ and $GG$ contributions as a function
of $\sqrt{s}$ vary for each pion PDFs, reflecting the differences
among their parton distributions. For SMRS and GRV the $GG$
contribution starts to dominate the cross section around $\sqrt{s}=20$
GeV. For xFitter and JAM the corresponding values are larger at
$\sim$$\sqrt{s}=35$ GeV because of their relatively reduced gluon
strength in the valence region.

\begin{table}[htbp]   
\centering
\begin{tabular}{|c|c|c|c|c|}
\hline
 & SMRS & GRV & JAM & xFitter \\
\hline
 $\chi^2_{total}/\text{ndf}$  &   3.1 &   3.4 &   4.8 &   4.3 \\ 
\hline
 $\chi^2/\text{ndp}|^{\pi^-}_{\sigma(J/\psi)}$&   2.7 &   3.7 &   8.4 &   6.4 \\ 
 $\chi^2/\text{ndp}|^{\pi^-}_{\sigma(\psi(2S))}$ &   0.4 &   0.7 &   0.4 &   0.2 \\ 
\hline
 $\langle \mathcal{O}_{8}^{J/\psi}[^{3}S_{1}] \rangle$ & $(6.9 \pm 0.3) \times 10^{-2} $ & $(9.5 \pm 0.4) \times 10^{-2}$ & $(8.3 \pm 0.3) \times 10^{-2}$ & $(7.4 \pm 0.9) \times 10^{-2}$ \\ 
 $\Delta_8^{J/\psi}$ & $(2.5 \pm 0.2) \times 10^{-2}$ & $(1.8 \pm 0.2) \times 10^{-2}$ & $(2.0 \pm 0.2) \times 10^{-2}$ & $(2.2 \pm 0.2) \times 10^{-2}$ \\ 
 $\langle \mathcal{O}_{8}^{\psi(2S)}[^{3}S_{1}] \rangle$ & $(2.1 \pm 0.4) \times 10^{-2}$ & $(2.6 \pm 0.2) \times 10^{-2}$ & $(2.6 \pm 0.1) \times 10^{-2}$ & $(2.3 \pm 0.4) \times 10^{-2}$ \\ 
 $\Delta_8^{\psi(2S)}$ & $(1.7 \pm 1.0) \times 10^{-3}$ & $(4.0 \pm 6.2) \times 10^{-4}$ & $(3.7 \pm 3.4) \times 10^{-4}$ & $(0.9 \pm 6.0) \times 10^{-3}$ \\ 
\hline
\end{tabular}
\caption {The $\chi^2$ values for the entire data sets and the
  individual $J/\psi$ and $\psi(2S)$ data set from the NRQCD
  calculations. The best-fit LDMEs for each pion PDF are in units of
  $\rm{GeV}^3$.}
\label{tab:LDME_PDF}
\end{table}


Table~\ref{tab:LDME_PDF} lists the $\chi^2$ values of NRQCD
calculations for various data sets and best-fit CO LDMEs for each pion
PDF. We find that the $\chi^2$ of the pion-induced $J/\psi$ data
strongly correlates with the gluon density of pions over the valence
quark regions~\cite{Chang:2020rdy}. The $GG$ contributions are similar
for GRV and SMRS, while those for xFitter and JAM are 50-80\% smaller
due to their weaker gluon strength at $x=0.1-0.6$,
relative to GRV and SMRS~\cite{Chang:2020rdy}. The deficiency of
xFitter and JAM in the $GG$ contribution leads to an underestimation
of the NRQCD calculations against the data over $\sqrt{s}=15-25$ GeV ,
as shown in Fig.~\ref{fig_jpsi_PDF}.

Table~\ref{tab:LDME_PDF} shows that the dependence of the best-fit
LDMEs for $J/\psi$ and $\psi(2S)$ to the pion PDFs is rather mild. We
also checked that the overall reduced $\chi^2/\text{ndf}$ of data for each
pion PDF has a very small variation when the calculations are done
with the best-fit LDMEs obtained using a different set of pion PDFs.

\section{Summary}
\label{sec:summary}

We have analyzed $J/\psi$ and $\psi(2S)$ hadroproduction data in
fixed-target experiments within the framework of NRQCD. The previously
reported difficulty~\cite{Beneke:1996tk} of obtaining a consistent
description of proton and pion data by NRQCD, is resolved by allowing
the CO LDMEs for $J/\psi$ and $\psi(2S)$ to be determined by a
simultaneous fit to both proton- and pion-induced data. The
pion-induced data, especially at low energies, have a sizable $q
\bar{q}$ contribution and thus provide a strong constraint on the
$\langle \mathcal{O}_{8}^{H}[^{3}S_{1}] \rangle$ LDMEs. This
sensitivity is much reduced if the analysis is restricted to the
proton-induced data alone. Consequently, the best-fit $\langle
\mathcal{O}_{8}^{H}[^{3}S_{1}] \rangle$ values for both $J/\psi$ and
$\psi(2S)$ are found to be about $\sim$10 times larger than that same
values obtained using the collider data. In contrast, the best-fit
values for $ \Delta_8^{H}$, partly responsible for the $GG$
contribution, are compatible for $J/\psi$ but more than 10 times
smaller for $\psi(2S)$, compared to the results of
Ref.~\cite{Beneke:1996tk}.

The resulting LDMEs combined with four different pion PDFs were used
to compare the NRQCD calculation with the pion-induced $J/\psi$
production cross-sections. The pronounced differences between the
predicted individual quark-antiquark annihilation and gluon-gluon
fusion terms result from the different shapes and magnitudes of the
corresponding PDF parametrizations. When compared to the total cross
section, the SMRS or GRV PDFs still provide a slightly better
description of the data than JAM or xFitter, suggesting that the data
favor those PDFs with larger gluon contents at medium and large
$x$. All these results are in line with our earlier
findings~\cite{Chang:2020rdy} obtained with the CEM.

We note that there are the recent state-of-the-art NLO NRQCD
fits~\cite{Butenschoen:2010rq, Ma:2010yw, Butenschoen:2011yh,
  Ma:2010jj} to collider data. The $\langle
\mathcal{O}_{8}^{J/\psi}[^{3}S_{1}] \rangle$ LDMEs found in
Refs.~\cite{Butenschoen:2010rq, Butenschoen:2011yh} are substantially
smaller than our results. In the fixed-target domain considered here
the mean $p_T$ values are lower than the charmonium mass. In this
low-$p_T$ region the factorization assumption of NRQCD may no longer
be valid. Higher-order corrections or other additional contributions
could therefore spoil the universality of the LDMEs. It will be
interesting to perform similar studies, including the pion-induced
data of charmonium production with the full NLO NRQCD formalism and
make a comparison with the current results in the future.

In the near future, new measurements of Drell-Yan as well as $J/\psi$
data in $\pi^- N$ reactions will be available from the CERN
COMPASS~\cite{COMPASS} and AMBER~\cite{AMBER} experiments. While
future theoretical advances are needed to further reduce the model
dependence in describing the $J/\psi$ production, we emphasize the
importance of including the pion-induced $J/\psi$ data in future pion
PDF global analysis.

\section*{Acknowledgments}
\label{sec:acknowledgments}

We thank Nobuo Sato and Ivan Novikov for providing us with
LHAPDF6-compatible grid files of JAM and xFitter PDFs. This work was
supported in part by the U.S. National Science Foundation (Grant
PHY18-22502) and the Ministry of Science and Technology of Taiwan.



\end{document}